\documentclass[aps,amsfonts,nofootinbib]{revtex4}
\usepackage{epsfig}
\usepackage{graphicx}
\usepackage{amsmath}
\begin{document}
\newcommand{\bb}{\begin{equation}}
\newcommand{\ee}{\end{equation}}
\newcommand{\eqb}{\begin{eqnarray}}
\newcommand{\eqf}{\end{eqnarray}}
\def\ka{\sqrt{\omega^2-m^2}}
\def\qmin{q_-}
\def\qmas{q_+}
\def\km{{k_+}}
\def\kmin{{k_-}}
\def\ob{{\bar{\omega}}}
\def\wkb{{\mbox{\tiny{WKB}}}}
\def\trans{{\mbox{\tiny{Trans}}}}
\def\refl{{\mbox{\tiny{Reflect}}}}
\def\kab{{\sqrt{\ob^2-m^2}}}
\newcommand{\1}{\'{\i}}
\preprint{}
\title{ Axions and ``Light Shining Through a Wall'': A Detailed Theoretical Analysis}
\author{Stephen L. Adler}
\email{adler@ias.edu} \affiliation{Institute for Advanced Study,
Einstein Drive,  Princeton N. J. 08540, USA}
\author{J. Gamboa}
\email{jgamboa@usach.cl} \affiliation{Departamento de F\'{\i}sica,
Universidad de Santiago de Chile, Casilla 307, Santiago, Chile }
\author{J. L\'opez-Sarri\'on}
\email{justinux75@gmail.com} \affiliation{Department of Physics,
The City College of the CUNY, New York 10031, USA}
\author{F. M\'endez}
\email{fmendez@usach.cl} \affiliation{Departamento de F\'{\i}sica,
Universidad de Santiago de Chile, Casilla 307, Santiago, Chile }

\begin{abstract}

We give a detailed study of axion-photon and photon-axion
conversion amplitudes, which enter the analysis of ``light shining
through a wall'' experiments.  Several different calculational
methods are employed and compared, and in all cases we retain a
nonzero axion mass. To leading order, we 
 find that when the photon frequency
$\omega$ is very close to the axion mass $m$, there is a threshold
cusp which significantly enhances the photon to axion conversion
amplitude, by a factor $\omega/\sqrt{\omega^2-m^2}$ relative to the
corresponding  axion to photon conversion process.  When $m=0$,
the enhancement factor  reduces to unity and the results of
previous calculations are recovered.  Our calculations include an
exact wave matching analysis, which  shows how unitarity is
maintained near threshold at $\omega=m$, and a discussion of the
case when the magnetic field extends into the ``wall'' region.
\end{abstract}
\pacs{PACS numbers:12.60.-i,11.30.Cp}
\date{\today}
\maketitle

\section{Introduction}
The axion is a remarkable idea proposed at the end of the
seventies in order to solve  the strong CP problem \cite{PQ}. The
possible existence of the axion could help solve long-standing
cosmological problems and, therefore, searching for it  is an
important issue \cite{review}.

Following on the seminal  papers of Sikivie \cite{sikivie} which showed that axions can be detected through axion-photon conversion in a magnetic field, 
twenty years ago  Van Bibber et al. proposed a ``light shining
through a wall" (LSW)  experiment where photon regeneration could
be used as an indication of an axion-photon coupling
\cite{bibben}. In the Van Bibber et al.  setup  (see Figure 1),
${\bf B}$ is an external magnetic field and the wall is opaque for
photons while it is transparent for the very weakly interacting
axions.   An x-ray version of this experiment has recently been
proposed by Rabad\' an et al. \cite{rabad}, and  two recent
optical laser {LSW} experiments have reported preliminary results
\cite{robill}, \cite{osquar}.

Motivated by the current interest in {LSW} experiments and axion
detection, we give here a detailed examination of the theory of
photon-axion and axion-photon conversion processes.  Our aims are
in part to compare different calculational methods, and in part to
examine threshold effects that appear when the axion mass is not
set equal to zero.

In the LSW experiments one can distinguish seven zones, as
sketched the figure.   We are mainly interested in the transition
zones I-II-III and V-VI-VII, namely, vacuum-magnetic field-vacuum
regions where an incident photon gives rise to an exiting axion,
and an incident axion gives rise to a regenerated photon. Although
we will mainly focus on the theory connected with these regions,
we will also briefly discuss what happens when the magnetic field
enters the absorbing wall region.

The paper is organized as follows.  In Sec. II we give the
Lagrangian, equations of motion, and some elementary properties
following from the equations of motion, such as the continuity
conditions, free space kinematics, conserved probability current,
and unitarity relation.  In Sec. III  we derive the leading order
photon-axion and axion-photon conversion amplitudes by a Green
function method, and in Sec. IV we repeat this derivation by a
WKB/eikonal method, in both cases allowing the magnetic field to
have a general dependence on $z$, the coordinate along the axis of
the experiment.   We find that when the axion mass is taken into
account, the ratio of the photon-axion conversion amplitude to the
axion-photon conversion amplitude is $\omega/k$, with $\omega$ and
$k=\sqrt{\omega^2-m^2}$ respectively the photon and axion wave
numbers. Since the leading order axion-photon amplitude violates
unitarity near threshold at $\omega=m$, in Sec. V we do an all
orders calculation for the case of a piecewise constant magnetic
field. This  gives the shape of the photon-axion conversion
amplitude near threshold, and allows us to check that the
unitarity constraint is obeyed.  In Sec. VI we give further
results following from the calculation of Sec. V.  We show that
the ratio of photon-axion to axion-photon amplitudes derived to
lowest order is in fact exact at all orders for a piecewise
constant magnetic field, we show that the all orders calculation
restores unitarity near threshold, and we give some rough
estimates of ``light through a wall'' and photon-axion conversion
probabilities integrated over the threshold cusp region. In Sec.
VII we briefly discuss what happens when the magnetic field
penetrates the wall region; although the wall strongly absorbs
photons, the weak coupling of photons to axions leads to a greatly
suppressed absorption of the axion wave. Finally, in Sec. VIII we
give a brief summary and suggestions for follow-up work.

\section{Lagrangian, equations of motion, and unitarity relations}

In this paper we will consider a system described by the action 
\bb {\cal S} = \int d^4x\,\left[ -\frac{1} 4 F^{\mu\nu} F_{\mu
\nu} - \frac{1}2 \phi\left(\partial^2 + m^2\right)\phi + \frac{g}
4 \phi F^{\mu\nu}{\tilde F}_{\mu \nu}\right]~~~, \label{1} \ee 
with ${\tilde F}^{\mu \nu}=\frac{1}{2}
\epsilon^{\mu\nu\alpha\beta}F_{\alpha\beta}$, where $\phi$ is a
real scalar field representing the axion, $F_{\mu \nu}$ is the
electromagnetic field tensor, and $g$ is the coupling constant.

The physical situation that we are interested in consists of
propagating photons and axions in the presence of a static
magnetic field as a background. We will suppose that this magnetic
background $B$ is pointing in the $x$ direction, so that the
relevant part of the interaction term is
\begin{equation}
\label{intterm}
 \int d^4x\, \beta \phi E_x~~~,
\end{equation}
where $\beta= g B$ and $E_x$ is the component of the photon
electric field parallel to the background magnetic field.

Neglecting components of the photon field that do not couple to
the axion through the magnetic field, and using Coulomb gauge,
(\ref{1}) becomes
\bb {\cal S} = -\frac{1}2 \int d^4x\, \left[a\partial^2 a
+\phi(\partial^2 + m^2)\phi - 2\beta \phi
\partial_t a \right]~~~, \ee
where $a$  is a real potential which determines the relevant
electric field component through $E_x= \partial_t a$.  From this
action we get the equations of motion,
\begin{eqnarray}
\label{eqmot}
(\partial^2  + m^2)\phi &=& \beta\partial_t a~~~, \nonumber\\
\partial^2 a & = & -\beta \partial_t\phi~~~,
\end{eqnarray}
so that, specializing to the case in which the propagation
direction is the $z$ axis, we have
\begin{eqnarray}
\label{zaxeqs}
(\partial_t^2-\partial_z^2  + m^2)\phi(z,t) &=& \beta(z)\partial_t
a(z,t)~~~, \nonumber\\
 (\partial_t^2-\partial_z^2) a(z,t) & = & -\beta(z)
\partial_t\phi(z,t)~~~.
\end{eqnarray}

The entire analysis of this paper will follow from the equations
of motion (\ref{zaxeqs}).  In particular, since these are second
order in $z$, the functions $\phi(z,t)$ and $a(z,t)$, and their
first spatial derivatives, must be continuous.  Since no complex
numbers enter the equations, we can follow the usual method of
generalizing the wave amplitudes from real to complex. Multiplying
the equation  for $\phi$ by $\phi^*$, multiplying the equation for
$\phi^*$ by $\phi$ and subtracting; similarly multiplying the
equation  for $a$ by $a^*$, multiplying the equation for $a^*$ by
$a$ and subtracting; and finally adding the two resultant
equations, one finds a current conservation equation
\begin{eqnarray}
\label{conseq}
\partial_t j_{\,0}+\partial_z j_z=0~~~, \nonumber\\
j_{\,0}=\phi^*\partial_t \phi-\partial_t\phi^* \phi +a^*
\partial_t
a-\partial_t a^* a +\beta(z) (a^*\phi-\phi^* a)~~~,  \nonumber\\
j_z=\partial_z\phi^*\phi-\phi^*\partial_z \phi +\partial_z a^* a
-a^* \partial_z a~~~.
\end{eqnarray}

Specializing now to the case of an incident wave with time
dependence $f(z)e^{-i\omega t}$, where $f$ denotes the fields
$\phi$ or $A$, $j_{\,0}$ becomes independent of time, and so
$\partial_t j_{\,0}=0$.  The current conservation equation then
becomes $\partial_z j_z=0$, which we use to give a useful
unitarity relation, as follows.  In free space regions where the
magnetic field vanishes (zones I, III, V, and VII of Figure 1),
the $z$ dependence of $A$ is $e^{\pm i \omega z}$ and of $\phi$ is
$e^{\pm i k z}$, where
\begin{eqnarray}
\label{kdef} k=\sqrt{\omega^2-m^2}~~~.
\end{eqnarray}
Assuming no incident waves from the right of the region where the
$B$ field is nonvanishing, the fields to the left will be given by
\begin{eqnarray}
\label{leftwave}
 \phi(z,t)=e^{-i\omega t}(\phi_I e^{ikz}+\phi_R
e^{-ikz})~~~,\nonumber\\
a(z,t)=e^{-i\omega t}(a_I e^{i\omega z}+a_R e^{-i\omega z})~~~,
\end{eqnarray}
and the fields to the right of the $B$ field region will be given
by
\begin{eqnarray}
\label{rightwave}
 \phi(z,t)=e^{-i\omega t}\phi_T e^{ikz}~~~,\nonumber\\
a(z,t)=e^{-i\omega t}a_T e^{i\omega z}~~~.
\end{eqnarray}
Substituting ({\ref{leftwave}) and (\ref{rightwave}) into
(\ref{conseq}) to give $j_z$ on the left and right, respectively,
and equating these, we get the unitarity relation between
incident, reflected, and transmitted wave amplitudes,
\begin{eqnarray}
\label{unitarity}
 k|\phi_I|^2 + \omega
 |a_I|^2=k(|\phi_R|^2+|\phi_T|^2)+\omega(|a_R|^2+|a_T|^2)~~~.
 \end{eqnarray}
 This equation will play a role in our subsequent discussion.

\section{Green function calculation of photon-axion and
axion-photon conversion}

We turn now to a calculation of the photon-axion and axion-photon
conversion amplitudes, to first order in the magnetic field
strength parameter $\beta(z)$.

By performing a perturbation expansion in powers of $\beta$ we
write the fields as follows
\begin{subequations}
\label{ex}
\begin{eqnarray}
\label{ex1} \phi(z,t)&=&\phi^{(0)}(z,t)+\phi^{(1)}(z,t)+\cdots,
\\
\label{ex2} a(z,t)&=&a^{(0)}(z,t)+a^{(1)}(z,t)+\cdots,
\end{eqnarray}
\end{subequations}
where
\begin{subequations}
\label{torr}
\begin{eqnarray}
\label{torr1} \left(\partial_t^2
-\partial_z^2+m^2\right)\phi^{(0)}(z,t)&=&0~~~,
\\
\label{torr2} \left(\partial_t^2
-\partial_z^2\right)a^{(0)}(z,t)&=&0~~~,
\\
\label{torr3} \left(\partial_t^2
-\partial_z^2+m^2\right)\phi^{(1)}(z,t)-\beta(z)\partial_t
a^{(0)}(z,t)&=&0~~~,
\\
\label{torr4} \left(\partial_t^2
-\partial_z^2\right)a^{(1)}(z,t)+\beta(z)\partial_t
\phi^{(0)}(z,t)&=&0~~~.
\end{eqnarray}
\end{subequations}

In the last two equations, unperturbed fields [solutions of
(\ref{torr1}) and (\ref{torr2})] are sources for perturbed ones,
so we will calculate the Green function for the differential
operator in (\ref{torr1}). Clearly, for case of the differential
operator in  (\ref{torr2}) it is enough to put $m=0$. Thus, we are
interested in the solutions $G_m(z,t)$ of
\begin{equation}
\label{green}
\left(\partial_t^2-\partial_z^2+m^2\right)G_m(z,t)=\delta(z)\delta(t)~~~.
\end{equation}
 A direct calculation in Fourier space shows that the desired Green function is
 \begin{equation}
 \label{expligreen}
 G_m(z,t)=-\frac{1}{(2\pi)^2}\int dkd\omega\frac{e^{i(kz-\omega
t)}}{(\omega-\omega_k)(\omega+\omega_k)}~~~,
 \end{equation}
with $\omega_k=\sqrt{k^2+m^2}$.

Since we want the retarded solution, that is $G_m(z,t)=0$ for
$t<0$, we integrate on the real $\omega$ axis, with poles $\mp
\omega_k$ shifted to $\mp \omega_k-i\epsilon$.  When $t<0$,
closing the contour up gives 0; when $t>0$, closing the contour
down gives the retarded Green function
\begin{equation}
\label{intgreen}
G_m^R(z,t)=-\theta(t)\left(\frac{i}{4\pi}\right)\int\frac{dk}{\omega_k
} \left(e^{i(kz+\omega_k t)}-e^{i(kz-\omega_k t)}
\right)\equiv-\theta(t)\Delta_m(z,t)~~~.
\end{equation}
Writing $\Delta_m$ in a more compact form we obtain
\begin{eqnarray}
\label{greenm} \Delta_m(z,t)=\int
\frac{dk}{2\pi}\,\frac{\sin(k|z|-\omega_kt) }{\omega_k}~~~,
\end{eqnarray}
and for the photon, taking $m=0$,  we find
\begin{eqnarray}
\label{green0}
G_0^R(z,t)=-\theta(t)\Delta_0(z,t)=\frac{1}{2}\theta(t-|z|)~~~.
\end{eqnarray}
Returning now to the solution of (\ref{torr3}), for the case
$\phi^{(0)}=0$ of no incident axion, we take the incident photon
field as a monochromatic wave  $ a^{(0)}=e^{i\bar{\omega}(z-t)}$,
so that the inhomogeneous equation (\ref{torr3}) reads
\begin{eqnarray}
\label{phi1eq}
\left(\partial_t^2-\partial_z^2+m^2\right)\phi^{(1)}(z, t)=-i\bar
{\omega}\beta(z)\,e^{i\ob(z-t)}~~~.
\end{eqnarray}
The solution to (\ref{phi1eq}), constructed using the Green
function (\ref{intgreen}) is
\begin{eqnarray}
\phi^{(1)}(z,t)&=&-i\ob\int_0^L dz'\beta(z') e^{i\ob z'}
\int_{-\infty}^{\infty} dt'G_m^R(z-z',t-t')e^{-i\ob t'} \,
\nonumber
\\
&=&-i\ob\,e^{-i\ob t}\int_0^L dz'\beta(z')e^{i\ob
z'}\int_{-\infty}^{\infty} \frac{dk}{\omega_k}
\left(\frac{-i}{4\pi}\right)e^{ik(z-z')}\nonumber
\\
&&\times\int_{-\infty}^\infty
dt'\,\theta(t-t')e^{i\ob(t-t')}\left(e^{
i(t-t')\omega_k}-e^{-i(t-t')\omega_k}\right)~~~,
\end{eqnarray}
where we have assumed that $\beta(z)=0$ outside the interval $0
\leq z \leq L$.

The integral on $t'$ can be done straightforwardly by defining
$T=t-t'$ and noting that, due to the $\theta$ function, there are
contributions only for $T>0$.  Then
\begin{eqnarray}
\phi^{(1)}(z,t)&=&-i\ob\,e^{-i\ob t}\int_0^Ldz'\beta(z')e^{i\ob
z'}\int_{-\infty}^{\infty} \frac{dk}{\omega_k}
\left(\frac{-i}{4\pi}\right)e^{ik(z-z')}\left(\frac{2i\omega_k}{\omega
_k^2- \ob^2}\right) \nonumber
\\
\label{tint} &=& -i\ob\,e^{-i\ob t}\int_0^Ldz'\beta(z')e^{i\ob
z'}\int_{-\infty}^{\infty}
\frac{dk}{2\pi}\,\frac{e^{ik(z-z')}}{(k-\sqrt{\ob^2-m^2})(k+\sqrt{\ob^2-m^2})}~~~.
\end{eqnarray}

The integral on $k$ can be performed as usual in the complex
plane.  The boundary condition that $a^{(0)}$ vanishes in the
infinite past can be included by giving $\ob$ a small positive
imaginary part $\ob \to \ob + i\epsilon,~\epsilon>0$, so that
\begin{eqnarray}
\label{contour} \pm \sqrt{(\ob+i\epsilon)^2-m^2}\sim \pm
\sqrt{\ob^2-m^2}\pm\frac{i\epsilon\ob}{\sqrt{\ob^2-m^2}} =\pm
\sqrt{\ob^2-m^2} \pm i \epsilon',~~\epsilon'>0~~~.
\end{eqnarray}
We see that for $z-z'>0$ the contour can be closed up, circling
the pole at $k=\sqrt{\ob^2-m^2}+i\epsilon'$, while for $z-z'<0$
the contour can be closed down, circling the pole at
$k=-\sqrt{\ob^2-m^2}-i\epsilon'$.  Equation (\ref{tint}) thus
gives
\begin{eqnarray}
\label{greenfinal1} \phi^{(1)}(z,t)&=&
\frac{\ob}{2\sqrt{\ob^2-m^2}}\bigg[ e^{-i\ob t+iz\kab }\int_0^L
dz' \theta(z-z')\beta(z')e^{iz'(\ob-\kab)} \nonumber
\\
&+& e^{-\ob t-iz\kab }\int_0^L dz'
\theta(z'-z)\beta(z')e^{iz'(\ob+\kab)} \bigg]\nonumber
\\
&=&\frac{\ob}{2\sqrt{\ob^2-m^2}}\bigg[ e^{-i\ob t+iz\kab }\int_0^z
dz' \beta(z')e^{iz'(\ob-\kab)}  \nonumber
\\
&+& e^{-i\ob t-iz\kab }\int_z^L dz'\beta(z')e^{iz'(\ob+\kab)}
\bigg]~~~.
\end{eqnarray}

We see that the axion amplitude has a transmitted  and a reflected
part,
\begin{eqnarray}
\label{ampsplit}
\phi^{(1)}(z,t)=\phi^{(1)}_\trans(z,t)~~+~~\phi^{(1)}_\refl(z,t)~~~,
\end{eqnarray}
with
\begin{eqnarray}
\label{greenfinal10}
\phi^{(1)}_\trans(z,t)&=&\frac{\ob}{2\sqrt{\ob^2-m^2}}\,\,
e^{-i\ob t+iz\kab }\int_0^z
dz'\beta(z')e^{iz'(\ob-\kab)}~~~,\nonumber
\\
\phi^{(1)}_\refl(z,t)&=&\frac{\ob}{2\sqrt{\ob^2-m^2}}\,\, e^{-i\ob
t-iz\kab }\int_z^L dz'\beta(z')e^{iz'(\ob+\kab)}~~~,
\end{eqnarray}
for general $z$, or for $z$ values outside the magnetic field
region,
\begin{eqnarray}
\phi^{(1)}_\trans(z,t)&=&\frac{\ob}{2\sqrt{\ob^2-m^2}}\,\,
e^{-i\ob t+iz\kab }\int_0^L dz'\beta(z')e^{iz'(\ob-\kab)}~,~~z\geq
 L~~~,\nonumber
\\
\label{greenfinal2}
\phi^{(1)}_\refl(z,t)&=&\frac{\ob}{2\sqrt{\ob^2-m^2}}\,\, e^{-i\ob
t-iz\kab }\int_0^L dz' \beta(z')e^{iz'(\ob+\kab)}~,~~z\leq 0~~~.
\end{eqnarray}

{}From (\ref{greenfinal2}) we see that if the exponential in the
integral for $\phi^{(1)}_\trans$ is rapidly varying on the scale
over which $\beta(z)$ varies, the axion transmission is strongly
suppressed.  This statement can be given a quantitative form by an
integration by parts,
\begin{equation}
\label{intparts} \int_0^L dz'\beta(z')e^{iz'(\ob-\kab)} =i\int_0^L
dz' \frac {\beta'(z')} {\ob-\kab}e^{iz'(\ob-\kab)}~~~,
\end{equation}
where we have used the fact that $\beta(0)=\beta(L)=0$ to drop the
surface terms.  If we assume that $|\beta'(z)/\beta(z)| < 1/h$,
with $h$ a measure of the characteristic distance over which the
magnetic field varies, and that $|\beta(z)|<\beta_M$, with
$\beta_M$ a measure of the maximum magnetic field, then
substitution of (\ref{intparts}) into (\ref{greenfinal2}) gives
the inequality
\begin{equation}
\label{ineq}
 |\phi^{(1)}_\trans| \leq \frac {\beta_M L } {2} \frac
{\ob} {\kab}
 \frac {1} {h(\ob-\kab)}~~~.
\end{equation}

Let us now consider $\beta(z)=\beta$, a constant. The previous
integrals can be done straightforwardly, and we obtain for the
transmitted part
\begin{eqnarray}
\label{transamp}
 \phi^{(1)}_\trans(z,t)=\frac{\beta
\ob\,e^{i(z\sqrt{\ob^2-m^2}-\ob t ) }}{m^2\sqrt{\ob^2-m^2}}
e^{\frac{iL}{2}(\ob-\sqrt{\ob^2-m^2})}
(\ob+\sqrt{\ob^2-m^2}\,)\sin\left(\frac{L}{2}(\ob-\sqrt{\ob^2-m^2}\,
)\right)~~~,
\end{eqnarray}
which has the modulus squared
\begin{eqnarray}
\label{transmodsq}
|\phi^{(1)}_\trans(z,t)|^2=\frac{\beta^2}{m^4}\left(\frac{\ob^2}{\ob^2-m^2}
\right) (\ob+\sqrt{\ob^2-m^2}\,)^2 \sin^2 \left(\frac { L } { 2 }
(\ob- \sqrt{\ob^2-m^2}\,)\right)~~~.
\end{eqnarray}
Let us recall now that the unitarity relation of (\ref{unitarity})
requires, for an incident photon amplitude of magnitude unity,
that the transmitted axion amplitude obey the unitarity constraint
$1 \geq (k/\omega) |\phi_T|^2$, which in the notation of this
section is $1 \geq (\sqrt{\ob^2-m^2}/\ob) |\phi^{(1)}(z \geq
L,t)|^2$.  From (\ref{transmodsq}) we see, however, that near
threshold at $\ob \simeq m$, we have
\begin{eqnarray}
\label{unitviol} (\sqrt{\ob^2-m^2}/\ob) |\phi^{(1)}(z \geq
L,t)|^2=\frac {\beta^2} {m \sqrt{\ob^2-m^2} } \sin^2 (mL/2)~~~,
\end{eqnarray}
which approaches infinity as $\ob \to m$.  Thus the lowest order
expression for the transmitted axion amplitude violates unitarity
near threshold; an all orders calculation, given below, is needed
to see how unitarity is restored, and to give the shape of the
axion transmission probability near threshold where
$k=\sqrt{\ob^2-m^2}\sim \beta^2/m$.

To complete the calculation, we will give the corresponding result
for the photon field amplitude induced by an axion wave incident
on the magnetic field region, with no incident photon ({\it i.e.},
$a^{(0)}=0$). Using the zero mass retarded Green function $G_0^R$
of (\ref{green0}) with the source term
 $-\beta(z)\partial_t\phi^{(0)}(z,t)=i\ob\beta(z)e^{i(\sqrt{\ob^2-m^2}z-\ob t)}$  in (\ref{torr4}), we
have
\begin{eqnarray}
a^{(1)}(z,t)&=&\int_0 ^L dz'\int_{-\infty}^\infty dt'
G_0^R(z-z',t-t')i\ob\beta(z')e^{i(\sqrt{\ob^2-m^2}z'-\ob t')}
\nonumber
\\
&=&\int_0 ^L dz'\int_{-\infty}^\infty dt'
\frac{1}{2}\theta(t-t'-|z-z'|)~
i\ob\beta(z')e^{i(\sqrt{\ob^2-m^2}z'-\ob t')}\nonumber
\\
\label{photamp1}
 &=&\frac{i\ob\,e^{-i\ob t}}{2}\int_0 ^L dz'
\beta(z')e^{iz'\sqrt{\ob^2-m^2}}\int_{|z-z'|}^\infty
 e^{i\ob T}\,dT~~~.
\end{eqnarray}
Finally, performing the integral on $T=t-t'$, we find for the
transmitted photon amplitude at $z \geq L$,
\begin{equation}
\label{greenphoton}
a^{(1)}_\trans(z,t)=-\frac{e^{i\ob(z-t)}}{2}~\int_0^L dz'
\beta(z')e^{iz'(\sqrt{\ob^2-m^2}-\ob)}~~~.
\end{equation}
We see that when the axion mass is nonzero, the ratio of
transmitted wave amplitude magnitudes is
$|\phi^{(1)}_\trans/a^{(1)}_\trans|=\ob/\sqrt{\ob^2-m^2}>1$. When
the axion mass $m$ vanishes, the transmitted amplitudes
(\ref{transamp}) and (\ref{greenphoton}) are identical up to a
sign, as analyzed in greater detail by Guendelman \cite{Guendel}
using his demonstration of an of an axion-photon duality.

\section{WKB/eikonal calculation of photon-axion and
axion-photon conversion}

In this section we calculate the photon-axion and axion-photon
conversion amplitudes, using a WKB/eikonal method. As an
alternative method of solving the equations of motion, with
$e^{-i\omega t}$ time dependence, we consider the following Ansatz
\begin{equation}
\label{solwkb} \phi(z,t)=\phi_0\,e^{i\theta(z)-i\omega
t},~~~~~~~~~ a(z,t)=a_0\,\,e^{i\chi(z)-i\omega t}~~~.
\end{equation}
The equations of motion (\ref{zaxeqs}) now read
\begin{eqnarray}
\label{eomwkb}
 \big(\omega^2-m^2-(\theta')^2+i\theta''\big)\phi-i\beta(z)\omega\,
a&=&0~~~,\nonumber
\\
 \big(\omega^2-{(\chi')}^2
 +i\chi''\big) a+i\beta(z)\omega\phi&=&0~~~.
\end{eqnarray}
Expanding  $\theta(z)$ and $\chi(z)$ in powers of $\beta$,
\begin{equation}
\label{exp}
 \theta(z)=\sqrt{\omega^2-m^2}z+\theta^{(1)}(z)+\cdots,
~~~~~~~~~ \chi(z)=\omega z+\chi^{(1)}(z)+\cdots,
\end{equation}
where the index $(1)$ indicates the first order term in the
expansion in  $\beta$, equations (\ref{eomwkb}) become
\begin{equation}
\label{orderone} \left(
\begin{array}{ccc}
2\sqrt{\omega^2-m^2}\,(\theta^{(1)})'\,e^{iz\sqrt{\omega^2-m^2}} &
& i\beta(z)\,\omega\,e^{iz\omega }
\\
-i\beta(z)\,\omega\,e^{iz\sqrt{\omega^2-m^2}} &  & 2\omega
(\chi^{(1)})'\, e^{iz\omega}
\end{array}\right)
\left(
\begin{array}{c}
\phi_0
\\
a_0
\end{array}\right)=0~~~,
\end{equation}
where for the moment we have dropped  second derivatives of phases
(as indicated below by a subscript WKB), and only first order
contributions in $\beta$ are considered.

Non-trivial solutions satisfy the null determinant condition
\begin{equation}
\label{nulldet}
(\theta^{(1)})'\,(\chi^{(1)})'=\frac{\omega\beta^2(z)}{4\sqrt{\omega^2
-m^2} }~~~.
\end{equation}
{}From the second row of (\ref{orderone}) we get
\begin{equation}
\label{chi1} \chi^{(1)}(z)=\frac{i\phi_0}{2
a_0}\int^{z}dz'\beta(z')\, e^{iz'(\sqrt{\omega^2-m^2}
-\omega)}~~~,
\end{equation}
while from the first row of (\ref{orderone}) (or using the null
determinant condition) we get
\begin{equation}
\label{teta1} \theta^{(1)}(z)=\frac{-ia_0}{2
\phi_0}\frac{\omega}{\sqrt{\omega^2-m^2}}\int^{z}dz'\beta(z')\,
e^{iz'(\omega-\sqrt{\omega^2-m^2})}~~~.
\end{equation}
This integral is the same as the one appearing in the transmitted
axion wave in  (\ref{greenfinal10}). In fact, we can  now
explicitly  write the axion field to first order in $\beta$, in
the approximation of neglecting second derivatives in
(\ref{orderone}), as
\begin{eqnarray}
\label{axwkb} & & \phi_\wkb(z,t)\simeq \phi_0
e^{i(z\sqrt{\omega^2-m^2}-\omega
t)}\big(1+i\theta^{(1)}(z)\big)\nonumber\\
&\simeq&\phi_0e^{i(z\sqrt{\omega^2-m^2}-\omega t)}\left(
1+\frac{a_0}{2
\phi_0}\frac{\omega}{\sqrt{\omega^2-m^2}}\int^{z}dz'\beta(z')\,
e^{iz'(\omega-\sqrt{\omega^2-m^2})} \right)~~~,
\end{eqnarray}
or an equivalent expression,
\begin{equation}
\label{fifinal}
\phi_\wkb(z,t)-\phi^{(0)}(z,t)\simeq~\frac{a_0}{2}\frac{\omega}{\sqrt{
\omega^2-m^2}} ~~e^{i(z\sqrt{\omega^2-m^2}-\omega
t)}~\int^{z}dz'\beta(z')\, e^{iz'(\omega-\sqrt{\omega^2-m^2})}~~~.
\end{equation}

The photon field satisfies a similar relation, namely
\begin{equation}
\label{afinal}
a_\wkb(z,t)-a^{(0)}(z,t)\simeq~-\frac{\phi_0}{2}~e^{i\omega(z-t)}~
\int^zdz'\beta(z') e^{iz'(\sqrt{\omega^2-m^2}-\omega)}~~~.
\end{equation}
With the substitution $\omega \to \ob$, and taking the lower limit
of integration as $z=0$ for the case when $\beta(z)$ vanishes for
$z\leq 0$, these agree with the transmission amplitudes obtained
by the Green function method.

Finally, let us perform the calculation, again up to first order
in $\beta$, but preserving the second derivatives of phases, so
that the exact first order amplitude is obtained, including the
reflection amplitudes calculated by the Green function method.
{}From (\ref{eomwkb}) we obtain
\begin{eqnarray}
\label{secphi}
\phi_0\,(\theta^{(1)})^{''}+2i\phi_0\,\sqrt{\omega^2-m^2}\,
(\theta^{(1)})'&=& a_0
\,\omega\,\beta(z)\,e^{i(\omega-\sqrt{\omega^2-m^2})z}~~~,
\end{eqnarray}
which implies
\begin{eqnarray}
\label{secphi1}
 \left( \phi_0
(\theta^{(1)})'\,e^{2iz\sqrt{\omega^2-m^2}}\right)'&=&
a_0\,\omega\,\beta(z)\,e^{i(\omega+\sqrt{\omega^2-m^2})z}~~~,
\end{eqnarray}
with the integral
\begin{eqnarray}
\label{secphiint} i \phi_0 (\theta^{(1)})'  &=&iC\,e^{-2iz
\sqrt{\omega^2-m^2}}+ ia_0\,\omega\,e^{-2iz
\sqrt{\omega^2-m^2}}\,\int_0^z dz'\beta(z') \,
e^{i(\omega+\sqrt{\omega^2-m^2})z'} ~~~.
\end{eqnarray}
Integrating again,  the phase $\theta^{(1)}$ is given by
\begin{eqnarray}
\label{tetfinal}
i\,\phi_0\,\theta^{(1)}&=&\tilde{C}+\frac{iC}{\sqrt{\omega^2-m^2}}
 \,e^{-iz\sqrt{\omega^2-m^2}}\sin\left(z\sqrt{\omega^2-m^2}
\right)\nonumber
\\
&&+ia_0\,\omega\,\int_0^zdz'\,e^{-2iz'
\sqrt{\omega^2-m^2}}\,\int_0^{z'}dz'' \beta(z'') \,
e^{i(\omega+\sqrt{\omega^2-m^2})z''}\,~~~.
\end{eqnarray}
Rewriting the final term by an integration by parts, we get
\begin{eqnarray}
\label{tetafindos}
i\,\phi_0\,\theta^{(1)}&=&\tilde{C}+\frac{iC}{\sqrt{\omega^2-m^2}}
 \,e^{-iz\sqrt{\omega^2-m^2}}\sin\left(z\sqrt{\omega^2-m^2}
\right)\nonumber
\\
&&-\frac{a_0\omega}{2\sqrt{\omega^2-m^2}}e^{-2iz\sqrt{\omega^2-m^2}}
\int_0^zdz'\beta(z')e^{iz'(\omega+\sqrt{\omega^2-m^2})} \nonumber
\\
&&+\frac{a_0\omega}{2\sqrt{\omega^2-m^2}}
\int_0^zdz'\beta(z')e^{iz'(\omega-\sqrt{\omega^2-m^2})}~~~.
\end{eqnarray}
Note that the last integral in (\ref{tetafindos}) is what we found
in (\ref{teta1}), where second derivatives were not considered.

The integration constants can now be determined  by the boundary
conditions, and in order to do this, let us write the complete
solution for the axion amplitude,
\begin{equation}
\label{axsoln}
 \phi(z,t)=e^{-i\omega
t+iz\ka}\left(\phi_0+i\phi_0\theta^{(1)} +\cdots\right)~~~.
\end{equation}
Substituting (\ref{tetafindos})  and writing explicitly the left
and right movers, we have
\begin{eqnarray}
\label{previous} \phi(z,t)&=&e^{i(z\ka-\omega
t)}\left[\phi_0+\tilde{C}+\frac{C} {2\ka}
+\frac{a_0\omega}{2\ka}\int_0^zdz'\beta(z')e^{iz'(\omega-\ka)}
\right]\nonumber
\\
&-&e^{-i(z\ka+\omega t)}\left[\frac{C}{2\ka} +
\frac{a_0\omega}{2\ka} \int_0^z dz'\beta(z')e^{iz'(\omega+\ka)}
\right]~~~.
\end{eqnarray}

Since we do not have incident (or right moving) axions at $z=0$,
the following condition must be satisfied
\begin{equation}
\label{zzero}
 \phi_0+\tilde{C}+\frac{C}{2\ka} =0~~~,
\end{equation}
while for  $z>L$, there must be only right moving axions, and
therefore
\begin{equation}
\label{zell}
 C + a_0\omega
\int_0^Ldz'\beta(z')e^{iz'(\omega+\ka)}=0~~~.
\end{equation}
Solving (\ref{zzero}) and (\ref{zell}) for the integration
constants $C,\,\tilde C$, and substituting back into
(\ref{tetafindos}), we obtain finally
\begin{eqnarray}
\label{finalwkb}
\phi(z,t)&=&e^{i(z\ka-\omega
t)}\,\,\frac{a_0\omega}{2\ka}
\int_0^zdz'\beta(z')e^{iz'(\omega-\ka)} \nonumber
\\
&+&e^{-i(z\ka+\omega t)}\,\frac{a_0\omega}{2\ka}\,
\int_z^Ldz'\beta(z')e^{iz'(\omega+\ka)} ~~~,
\end{eqnarray}
in agreement with the transmitted and reflected waves obtained in
(\ref{greenfinal10}) by the Green function method.

When the axion mass is very small, so that $m<<\omega$, and
$\beta(z)$ is smooth, the reflected axion wave amplitude is much
smaller than the transmitted axion wave amplitude, because of the
rapid oscillation of the phase factor $e^{iz'(\omega +\ka)}$.
This small amplitude is what is neglected in making the WKB
approximation of neglecting second derivatives of the phases
$\theta,~\chi$ in going from (\ref{eomwkb}) to (\ref{orderone}).

\section{All orders calculation: details}

We turn in this section to an all orders calculation of
photon-axion and axion-photon conversion, for the case of a
magnetic field that vanishes for $z<0$ and $z>L$, and is a
constant $B$, so that $gB=\beta$ is constant, in the interval
$0<z<L$.  We assume a time dependence $e^{-i\omega t}$ throughout,
so that the equations of motion have solutions of the form
$f(z)e^{-i\omega t}$, where $f$ denotes the fields $\phi$ or $a$;
henceforth in this section, the time-dependence factor
$e^{-i\omega t}$ will not be shown explicitly.

Referring back to (\ref{leftwave}), the photon and axion waves in
the region $z\leq 0$ have the form
\begin{eqnarray}
\label{leftwave1}
 \phi(z)=\phi_I e^{ikz}+\phi_Re^{-ikz}~~~,\nonumber\\
a(z)=a_I e^{i\omega z}+a_R e^{-i\omega z}~~~.
\end{eqnarray}
Generalizing (\ref{rightwave}) to allow incoming photon and axion
waves from the right (which will be equated to zero later in the
calculation), the photon and axion waves in the region $z\geq L$
have the form
\begin{eqnarray}
\label{rightwave1}
 \phi(z)=\phi_T e^{ikz} + \tilde \phi e^{-ikz}~~~,\nonumber\\
a(z)=a_T e^{i\omega z} + \tilde a e^{-i\omega z}~~~.
\end{eqnarray}

In the region $0\leq z \leq L$, the propagation eigenmodes are
obtained by solving the coupled equations (\ref{zaxeqs}), which
with the assumed $e^{-i\omega t}$ time dependence take the form
\begin{eqnarray}
\label{zaxeqs1}
(-\omega^2-\partial_z^2  + m^2)\phi(z) &=&
-i\omega \beta a(z)~~~, \nonumber\\
 (-\omega^2-\partial_z^2) a(z) & = & i\omega \beta \phi(z)~~~.
\end{eqnarray}
Assuming an $e^{iKz}$ dependence of both $\phi(z)$  and $a(z)$, by
substituting $\phi(z)=\phi e^{iKz}$~, $a(z)=ae^{iKz}$, we find the
coupled eigenmode equations
\begin{eqnarray}
\label{modeqs}
(-\omega^2+K^2  + m^2)\phi &=&
-i\omega \beta a, \nonumber\\
 (-\omega^2+K^2) a & = & i\omega \beta \phi~~~,
\end{eqnarray}
which require $K$ to obey the quartic equation
\begin{equation}
\label{Keq} (-\omega^2+K^2+m^2)(-\omega^2+K^2)=\omega^2\beta^2~~~.
\end{equation}
Solving this, we find that the fields propagate with four possible
values for $K$, either $K=\pm k_+$ or $K=\pm k_-$, with
\begin{equation}
\label{Kvalues}
(k_+)^2=\omega^2-\frac{m^2}{2}\left(1-\sqrt{1+x^2}\right)~,~~~
(k_-)^2=\omega^2-\frac{m^2}{2}\left(1+\sqrt{1+x^2}\right)~~~,
\end{equation}
where
\begin{equation}
\label{xdef}
 x=2\beta\omega /m^2~~~.
\end{equation}
For zero magnetic field, we have $x=0$ and then $\km$ gives the
photon mode,  and $\kmin$ gives the axion mode. The most general
solution in the magnetic field region is
\begin{eqnarray}
\label{solax}
\phi(z)&=&\Phi_0^+~e^{iz\km}+\Phi_0^-e^{iz\kmin}+{\varphi}_0^+~e^
{ -iz\km } + {\varphi}_0^-~e^{ -iz\kmin }~~~,
\\
\label{solfo} a(z)&=&A_0^+~e^{iz\km}+A_0^-e^{iz\kmin}+{a}_0^+~e^ {
-iz\km } + {a}_0^-~e^{ -iz\kmin }~~~.
\end{eqnarray}
There are restrictions on the integration constants coming from
the equations of motion (\ref{modeqs});  they are
\begin{equation}
\label{inconst}
 \Phi_0^{+}=\delta~A_0^{+},~~~
\varphi_0^{+}=\delta~a_0^{+},~~~ A_0^{-}=\delta~\Phi_0^{-},~~~
a_0^{-}=\delta~\varphi_0^{-}~~~,
\end{equation}
with
\begin{eqnarray}
\label{deltadef} \delta=\frac{-ix}{1+\sqrt{1+x^2}}~~~.
\end{eqnarray}
Thus, the solutions in the magnetic field region take the form
\begin{eqnarray}
\label{midax}
\phi(z)&=&\delta~A_0^+~e^{iz\km}+\Phi_0^-e^{iz\kmin}+\delta
 a_0^+~e^{ -iz\km } + {\varphi}_0^-~e^{ -iz\kmin }~~~,
\\
\label{midpho}
a(z)&=&A_0^+~e^{iz\km}+\delta~\Phi_0^-e^{iz\kmin}+{a}_0^+~e^ {
-iz\km } + \delta~{\varphi}_0^-~e^{ -iz\kmin }~~~.
\end{eqnarray}
For $x\to 0$ we see that $\delta\to 0$, so that $\km \to \omega$
and $\kmin \to k$, and we recover free photons and axions. An
important property of the above equations is that when the sign of
the square root $\sqrt{1+x^2}$ is changed from $+$ to $-$, the
modes $\km$ and $\kmin$ are interchanged, that is, $\km
\leftrightarrow \kmin$, and also $\delta$ is replaced by
$\frac{-ix}{1-\sqrt{1+x^2}}= \frac{1+\sqrt{1+x^2}}{-ix} =
1/\delta$, that is $\delta \leftrightarrow 1/\delta$.  This
property will be used in Appendix B  to show that the ${
S}$-matrix elements governing all physical effects are independent
of the branch choice of the square root (a behavior similar to
that found in the study of vacuum birefringence effects
\cite{adler,bial,toll} when one does a wave-matching calculation
\cite{adler1,biswas} for the case of a rotating magnetic field).

We now impose continuity conditions on fields and spatial
derivatives at $z=0$, where  $\beta$ changes from zero to nonzero,
and at $z=L$, where $\beta$ changes back to zero.

\subsection{Continuity conditions at  $z=0$}

At $z=0$, the continuity conditions for the fields are
\begin{eqnarray}
\label{axcont} \phi_I+\phi_R&=&\delta~A_0^++\Phi_0^-+\delta
 a_0^+~ + {\varphi}_0^-~~~,
\\
\label{fotcont} a_I +a_R &=& A_0^++\delta~\Phi_0^-+{a}_0^++
\delta~{\varphi}_0^-~~~,
\end{eqnarray}
while the continuity conditions for first spatial derivatives are
\begin{eqnarray}
\label{deraxcont} k(\phi_I-\phi_R)&=&\delta~k_+(A_0^+-a_0^+)
+k_-(\Phi_0^- - {\varphi}_0^-)~~~,
\\
\label{derfotcont} \omega(a_I -a_R) &=&
k_+~(A_0^+-{a}_0^+)+\delta~k_-(\Phi_0^-- {\varphi}_0^-)~~~.
\end{eqnarray}

This set of four equations can be recast in matrix form as
\begin{equation}
\label{matcond1}  M_1~\left(
\begin{array}{c}
\phi_I \\ \phi_R \\a_I \\ a_R
\end{array}\right) = M_2~\left(
\begin{array}{c}
\Phi_0^- \\ \varphi_0^- \\ A_0^+ \\ a_0^+
\end{array}\right)~~~,
\end{equation}
where the matrices $M_1$ and $M_2$ can be read off from
(\ref{axcont}-\ref{derfotcont}), and are given in Appendix A. It
is straightforward to invert the matrix $M_1$, and defining
$M_{12}=M_1^{-1}M_2$, (\ref{matcond1}) can be rewritten as
\begin{equation}
\label{matcond2}  \left(
\begin{array}{c}
\phi_I \\ \phi_R \\a_I \\ a_R
\end{array}\right) = M_{12}~\left(
\begin{array}{c}
\Phi_0^- \\ \varphi_0^- \\ A_0^+ \\ a_0^+
\end{array}\right)~~~,
\end{equation}
with the matrix $M_{12}$ given in Appendix A.

\subsection{Continuity conditions at $z=L$}

At $z=L$, the continuity conditions for the fields are
\begin{eqnarray}
\label{contaxL} \delta~A_0^+~e^{iL\km}+\Phi_0^-e^{iL\kmin}+\delta
 a_0^+~e^{ -iL\km } + {\varphi}_0^-~e^{ -iL\kmin }&=&
\phi_T~e^{ikL}+\tilde \phi ~e^{-ikL}~~~,\nonumber
\\
&&
\\
\label{contfoL}
A_0^+~e^{iL\km}+\delta~\Phi_0^-e^{iL\kmin}+{a}_0^+~e^ { -iL\km } +
\delta~{\varphi}_0^-~e^{ -iL\kmin }&=&a_T~e^{i\omega L}+\tilde
a~e^{-i\omega L}~~~,
\end{eqnarray}
while the continuity conditions for first spatial derivatives are
\begin{eqnarray}
\label{dcontaxL} \delta~k_+(A_0^+~e^{iL\km}-
 a_0^+~e^{ -iL\km })+k_-(\Phi_0^-e^{iL\kmin} - {\varphi}_0^-~e^{
-iL\kmin })&=& k(\phi_T~e^{ikL}- \tilde
\phi~e^{-ikL})~~~,\nonumber
\\
&&
\\
\label{dcontfoL} k_+(A_0^+~e^{iL\km}-{a}_0^+~e^ { -iL\km }
)+\delta~k_-(\Phi_0^-e^{iL\kmin}- {\varphi}_0^-~e^{ -iL\kmin
})&=&\omega(a_T~e^{i\omega L}-\tilde a~e^{-i\omega L})~~~.
\end{eqnarray}

Again, this  set of four equations can be recast in matrix form as
\begin{equation}
\label{matccond3} M_3~\left(
\begin{array}{c}
\Phi_0^- \\ \varphi_0^- \\ A_0^+ \\ a_0^+
\end{array}\right) = M_4~\left(
\begin{array}{c}
\phi_T\\ \tilde\phi \\ a_T \\ \tilde a
\end{array}\right)~~~,
\end{equation}
where the matrices $M_3$ and $M_4$ can be read off from
(\ref{contaxL}-\ref{dcontfoL}),  and are given in Appendix A. And
again, it is straightforward to invert the matrix $M_3$, and
defining $M_{34}=M_3^{-1}M_4$, (\ref{matccond3}) can be rewritten
as
\begin{equation}
\label{matccond4} \left(
\begin{array}{c}
\Phi_0^- \\ \varphi_0^- \\ A_0^+ \\ a_0^+
\end{array}\right) = M_{34}~\left(
\begin{array}{c}
\phi_T\\ \tilde\phi \\ a_T \\ \tilde a
\end{array}\right)~~~,
\end{equation}
with the matrix $M_{34}$ given in Appendix A.

\subsection{Construction of the $S$ matrix}

Combining (\ref{matcond2}) and (\ref{matccond4}), we get a matrix
relation between the incoming and outgoing wave amplitudes,

\begin{eqnarray}
\label{smatrix} \left(
\begin{array}{c}
\phi_I \\ \phi_R \\a_I \\ a_R
\end{array}\right)
&=& M_{12}~M_{34}~\left(
\begin{array}{c}
\phi_T\\ \tilde\phi \\ a_T \\ \tilde a
\end{array}\right)\nonumber
\\
&\equiv& {S}~\left(
\begin{array}{c}
\phi_T\\ \tilde\phi \\ a_T \\ \tilde a
\end{array}\right)~~~.
\end{eqnarray}

{}From the matrices $M_{12}$ and $M_{34}$ given in Appendix A,
explicit calculation gives the ${ S}$ matrix elements tabulated in
Appendix B.  We use them to solve the equation
 (\ref{smatrix}) for the following three cases.

\subsection{Incident photon only}

We assume a photon of unit amplitude incident from the left of
$z=0$, so that $a_I=1$, but no  axion incident from the left, so
that $\phi_I=0$, and no photon or axion incident from the right of
$z=L$, so that $\tilde a=\tilde \phi=0$.

With this configuration, the set of four equations that must be
solved is

\begin{eqnarray}
\label{incphoteqs}
 0&=&S_{11}\phi_T+S_{13}a_T~~~,\nonumber
\\
\phi_R&=&S_{21}\phi_T+S_{23}a_T~~~,\nonumber
\\
1&=&S_{31}\phi_T +S_{33}a_T~~~,\nonumber
\\
a_R&=&S_{41}\phi_T+S_{43}a_T~~~,
\end{eqnarray}

from which we obtain the transmitted and reflected axion and
photon amplitudes $\phi_T$, $\phi_R$, $a_T$, $a_R$,
\begin{eqnarray}
\label{amplitudes1}
\phi_T&=&\frac{S_{13}}{S_{13}S_{31}-S_{11}S_{33}}~~~,\nonumber
\\
\phi_R&=&\frac{S_{21}S_{13}-S_{23}S_{11}}
{S_{13}S_{31}-S_{11}S_{33}}~~~,\nonumber
\\
a_T&=&\frac{-S_{11}}{S_{13}S_{31}-S_{11}S_{33}}~~~,\nonumber
\\
a_R&=&\frac{S_{41}S_{13}-S_{43}S_{11}}{S_{13}S_{31}-S_{11}S_{33}}~~~.
\end{eqnarray}
Referring to (\ref{unitarity}), these amplitudes obey the
unitarity constraint
\begin{equation}
\label{unit1} 1=|a_R|^2+|a_T|^2 + \frac
{k}{\omega}(|\phi_R|^2+|\phi_T|^2)~~~.
\end{equation}

\subsection{Incident axion only}

We assume next an axion of unit amplitude incident from the left
of $z=0$, so that $\phi_I=1$, but no  photon incident from the
left, so that $a_I=0$, and again no photon or axion incident from
the right of $z=L$, so that $\tilde a=\tilde \phi=0$.

With this configuration, the set of four equations that must be
solved is
\begin{eqnarray}
\label{incaxeqs}
 1&=&S_{11}\phi_T+S_{13}a_T~~~,\nonumber
\\
\phi_R&=&S_{21}\phi_T+S_{23}a_T~~~,\nonumber
\\
0&=&S_{31}\phi_T +S_{33}a_T~~~,\nonumber
\\
a_R&=&S_{41}\phi_T+S_{43}a_T~~~,
\end{eqnarray}
from which we obtain the transmitted and reflected axion and
photon amplitudes $\phi_T$, $\phi_R$, $a_T$, $a_R$,
\begin{eqnarray}
\label{amplitudes2}
\phi_T&=&\frac{-S_{33}}{S_{13}S_{31}-S_{11}S_{33}}~~~,\nonumber
\\
\phi_R&=&\frac{S_{23}S_{31}-S_{21}S_{33}}
{S_{13}S_{31}-S_{11}S_{33}}~~~,\nonumber
\\
a_T&=&\frac{S_{31}}{S_{13}S_{31}-S_{11}S_{33}}~~~,\nonumber
\\
a_R&=&\frac{S_{43}S_{31}-S_{41}S_{33}}{S_{13}S_{31}-S_{11}S_{33}}~~~.
\end{eqnarray}
Referring again to (\ref{unitarity}), these amplitudes obey the
unitarity constraint
\begin{equation}
\label{unit2} 1=|\phi_R|^2+|\phi_T|^2 + \frac
{\omega}{k}(|a_R|^2+|a_T|^2)~~~.
\end{equation}

\subsection{Incident photon and axion}

We assume finally the more general case in which an  axion of
amplitude $\phi_I$ and a photon of amplitude $a_I$ are both
incident from the left of $z=0$,  but again no photon or axion are
incident from the right of $z=L$, so that $\tilde a=\tilde
\phi=0$.

In this case,  the set of four equations that must be solved is
\begin{eqnarray}
\label{incbotheqs}
 \phi_I&=&S_{11}\phi_T+S_{13}a_T~~~,\nonumber
\\
\phi_R&=&S_{21}\phi_T+S_{23}a_T~~~,\nonumber
\\
a_I&=&S_{31}\phi_T +S_{33}a_T~~~,\nonumber
\\
a_R&=&S_{41}\phi_T+S_{43}a_T~~~,
\end{eqnarray}
from which we obtain the transmitted and reflected axion and
photon amplitudes $\phi_T$, $\phi_R$, $a_T$, $a_R$,
\begin{eqnarray}
\label{amplitudes3}
\phi_T&=&\frac{-S_{33}\phi_I+S_{13}a_I}{S_{13}S_{31}-S_{11}S_{33}}~~~,\nonumber
\\
\phi_R&=&\frac{(S_{23}S_{31}-S_{21}S_{33})\phi_I+(S_{21}S_{13}-S_{23}S_{11})a_I}
{S_{13}S_{31}-S_{11}S_{33}}~~~,\nonumber
\\
a_T&=&\frac{S_{31}\phi_I-S_{11}a_I}{S_{13}S_{31}-S_{11}S_{33}}~~~,\nonumber
\\
a_R&=&\frac{(S_{43}S_{31}-S_{41}S_{33})\phi_I+(S_{41}S_{13}-S_{43}S_{11})a_I}
{S_{13}S_{31}-S_{11}S_{33}}~~~.
\end{eqnarray}
This is of course just $a_I$ times the amplitudes of
(\ref{amplitudes1}) plus $\phi_I$ times the amplitudes of
(\ref{amplitudes2}), as  expected from linearity.  The general
unitarity constraint of  (\ref{unitarity}) now gives, in addition
to the constraints on the $S$-matrix elements of (\ref{unit1}) and
(\ref{unit2}), the additional constraint
\begin{eqnarray}
\label{unit3}
 0&=&k {\rm
 Re}[(S_{23}S_{31}-S_{21}S_{33})(S_{21}S_{13}-S_{23}S_{11})^*-S_{33}S_{13}^*]\nonumber
 \\
 &+& \omega {\rm
 Re}[(S_{43}S_{31}-S_{41}S_{33})(S_{41}S_{13}-S_{43}S_{11})^*-S_{11}S_{31}^*]
 ~~~.
\end{eqnarray}

\section{All orders calculation: discussion}

We now explore various features of the all orders results obtained
in the previous section.  We begin by showing that to leading
order in the parameter $x=2\omega \beta/m^2$ introduced in
(\ref{xdef}), we recover the result of the Green function and
WKB/eikonal calculations for the transmitted axion amplitude. When
$x<<1$, expansion of (\ref{Kvalues}) and (\ref{deltadef}) gives
\begin{equation}
\label{expansion}
 k_-\simeq \sqrt{\omega^2-m^2}+{\cal O}(x^2),~~~~~
k_+\simeq\omega {\cal O}(x^2),~~~~~ \delta\sim-\frac{ix}{2}+{\cal
O}(x^3)~~~,
\end{equation}
from which we find
\begin{eqnarray}
\label{NandD}
S_{13}&=&\frac{ix}{4}\frac{\omega+\sqrt{\omega^2-m^2}}{\sqrt{\omega^2-m^2}}
[e^{iL(\omega-\sqrt{\omega^2-m^2})}-1] +{\cal O}(x^2)~~~,\nonumber
\\
S_{13}S_{31}&=&{\cal O}(x^2)~~~,\nonumber
\\
S_{11}&=&S_{33}=1+{\cal O }(x^2)~~~.\nonumber
\\
\end{eqnarray}
Substituting these into (\ref{amplitudes1}), we get for the
transmitted axion amplitude when a photon of unit amplitude is
incident on the magnetic field region,
\begin{equation}
\label{phit1} \phi_T=\frac{ix}{4}\frac
{\omega+\sqrt{\omega^2-m^2}}
{\sqrt{\omega^2-m^2}}(1-e^{iL(\omega-\sqrt{\omega^2-m^2})})~~~,
\end{equation}
which can be rewritten  in the usual form given in
(\ref{transamp}),
\begin{equation}
\label{phit2} \phi_T=\frac {x}{2}
\frac{(\omega+\sqrt{\omega^2-m^2})}{\sqrt{\omega^2-m^2}}
e^{i\frac{L}{2}(\omega-\sqrt{\omega^2-m^2})}~\sin(
\frac{L}{2}(\omega-\sqrt {\omega^2-m^2}))~~~.
\end{equation}

We next examine the relationship between the two amplitudes that
enter into light through a wall experiments: the amplitude
$\phi_T$ for the transmitted axion, when only a photon of unit
amplitude is incident [see (\ref{amplitudes1})], and the amplitude
$a_T$ for the transmitted photon, when only an axion of unit
amplitude is incident [see (\ref{amplitudes2})].  We have
\begin{equation}
\label{ratio}
 \frac{\phi_T({\rm incident~ photon}) }{a_T({\rm
incident~ axion} )}=\frac{S_{13}}{S_{31}}
=-\frac{\omega}{k}e^{iL(\omega-k)}~~~~,
\end{equation}
where we have made use of (B10).  Hence the ratio of magnitudes is
$\omega/\sqrt{\omega^2-m^2}$ to all orders in $x$, for a piecewise
constant $B$, generalizing the result that we found earlier to
leading order.

Finally we turn to the behavior of the axion and photon
transmission amplitudes when $\omega$ is very close to threshold,
where we expect to find that the all orders calculation eliminates
the unitarity violation that we noted at leading order.  Let us
parameterize the kinematic variables of the problem in terms of a
dimensionless parameter $y$, which takes the value $-1$ at
threshold $\omega=m$, and $0$ at the $\omega$ value where $k_-$
vanishes.  According to (\ref{Kvalues}), for the product
$k_+^2k_-^2$ we have
\begin{equation}
\label{product}
k_+^2k_-^2=(\omega^2-m^2/2)^2-(m^2/2)^2(1+x^2)=\omega^2(\omega^2-m^2-\beta^2)~~~,
\end{equation}
and so $k_-$ vanishes at $\omega^2=m^2+\beta^2$.  (For values of
$\omega$ smaller than this, down to threshold at $\omega=m$, the
wave number  $k_-$ becomes imaginary.)  We thus
parameterize $\omega$ as follows,
\begin{equation}
\label{param} \omega^2=m^2+(1+y)\beta^2~~~,
\end{equation}
and assuming that $|\beta|/m<<1$, and that $y$ is of order unity,
we get the following approximations for the kinematic quantities
of interest,
\begin{equation}
\omega\simeq m+{\cal O}(\beta^2)~,~~k=\sqrt{1+y}|\beta|~,~~
k_+\simeq m+{\cal O}(\beta^2)~,~~k_-\simeq\sqrt{y}|\beta|~,~~
\delta\simeq -i\beta/m~~~.
\end{equation}
The $S$ matrix elements that are needed can now be approximated as
follows, working to leading order in $|\beta|/m$ and in
$\sqrt{1+y}$,
\begin{eqnarray}
S_{13}&\simeq&\frac{i
\beta}{2|\beta|\sqrt{1+y}}(e^{iLm}-1)~~~,\nonumber
\\
S_{31}&\simeq&-\frac{i\beta}{2m}(1-e^{-iLm})~~~,\nonumber
\\
S_{11}&\simeq&1-\frac{i|\beta|}{2m\sqrt{1+y}\,}[\sin(mL)-mL]~~~,\nonumber
\\
S_{33}&\simeq&1~~~.
\end{eqnarray}
Substituting these into (\ref{amplitudes1}), we find that
$|\phi_T|^2$ near threshold is given by
\begin{equation}
\label{cusp1}
 |\phi_T|^2\simeq \frac{\sin^2(\frac{1}{2}mL)}
{\left(\frac{|\beta|}{m} \sin^2(\frac{1}{2}mL) +
\sqrt{1+y}\right)^2+\frac{|\beta|^2}{4m^2}\big(\sin(mL)-mL\big)^2}~~~,
\end{equation}
or rewritten in terms of $k=|\beta|\sqrt{1+y}$~,
\begin{equation}
\label{cusp2}
 |\phi_T|^2\simeq \frac{\sin^2(\frac{1}{2}mL)}
{\left(\frac{|\beta|}{m} \sin^2(\frac{1}{2}mL) +
\frac{k}{|\beta|}\right)^2+\frac{|\beta|^2}{4m^2}\big(\sin(mL)-mL\big)^2}~~~.
\end{equation}
Note that once $y>>1$, or equivalently $k>>|\beta|^2/m$, these
equations reduce to
\begin{equation}
\label{reduct}
 |\phi_T|^2 \simeq
\frac{\beta^2}{\omega^2-m^2}\sin^2(\frac{1}{2}mL)~~~,
\end{equation}
in agreement with the leading order result (\ref{transmodsq} )
evaluated at $\ob \simeq m$.  Thus the leading order result is
still valid when $k/m<<1$, but requires the correction coming from
the all orders calculation when $k/m<<|\beta|^2/m^2$.

Unitarity requires that $(k/\omega)|\phi_T|^2 \simeq
(k/m)|\phi_T|^2$ should smaller from unity; from (\ref{cusp2}) we
find
\begin{equation}
\label{unitest}
 \frac{k}{m}|\phi_T|^2 \leq  \frac
{\frac{k}{m}\sin^2(\frac{1}{2}mL)} {\left(\frac{|\beta|}{m}
\sin^2(\frac{1}{2}mL) + \frac{k}{|\beta|}\right)^2}\leq
\frac{1}{4}~~~,
\end{equation}
where we have used the fact that for positive $s$ and $t$,
\begin{equation}
\label{ineq20} \frac{st}{(s+t)^2} =\frac{1}{4} \frac
{(s+t)^2-(s-t)^2}{(s+t)^2} \leq \frac{1}{4}~~~.
\end{equation}

Using the results of (\ref{ratio}) and (\ref{cusp2}), the squared
magnitude of the photon transmission amplitude, for a unit
incident axion amplitude, is given near threshold by
\begin{equation}
\label{cusp3}
 |a_T|^2\simeq \frac{\frac{k^2}{m^2}\sin^2(\frac{1}{2}mL)}
{\left(\frac{|\beta|}{m} \sin^2(\frac{1}{2}mL) +
\frac{k}{|\beta|}\right)^2+\frac{|\beta|^2}{4m^2}\big(\sin(mL)-mL\big)^2}~~~.
\end{equation}
Assuming no axion attenuation in the wall, the overall photon
transmission probability $P(\omega)$ in a light through a wall
experiment, very near threshold, is the product of $|\phi_T|^2$ of
(\ref{cusp2}) and $|a_T|^2$ of (\ref{cusp3}).  In the regime where
$\beta^2/m^2<<k/m<<1$, $P(\omega)$ is well approximated by
\begin{equation}
\label{Papprox}
 P(\omega)\simeq \frac{\beta^4
\sin^4(\frac{1}{2}mL)}{2 m^3 (\omega-m)}~~~.
\end{equation}
Defining  $\bar P(\omega)$ as the average of $P(\omega)$ over an
interval extending from $\omega=m$ to $\omega=m+\Delta$,
integration of (\ref{Papprox}) with a lower cutoff of $k_L^2 \sim
\beta^2/m^2$, or equivalently, $\omega_L-m \sim \beta^4/(2m^3)$,
we find
\begin{equation}
\label{Pbar} \bar P\equiv \frac {1}{\Delta} \int_{m}^{m+\Delta}
d\omega P(\omega) \simeq  \frac {1}{\Delta} \int_{\omega_L-
m}^{\Delta} d(\omega -m)P(\omega) \simeq \frac{\beta^4}{2\Delta
m^3}\sin^4(\frac{1}{2}mL) \log\left(\frac
{2m^3\Delta}{\beta^4}\right)~~~.
\end{equation}
In a similar fashion, if we define the emerging axion flux from
photon to axion conversion in a single $B$ field region, for unit
incident photon flux,  as $F(\omega)=(k/m)|\phi_T|^2$, then in the
regime where $\beta^2/m^2<<k/m<<1$, $F(\omega)$ is well
approximated by
\begin{equation}
\label{Fapprox} F(\omega) \simeq
\frac{\beta^2}{mk}\sin^2(\frac{1}{2}mL) \simeq \frac{\beta^2} {
m^{3/2}\sqrt{2(\omega-m)} }\sin^2(\frac{1}{2}mL) ~~~.
\end{equation}
Defining  $\bar F(\omega)$ as the average of $F(\omega)$ over an
interval extending from $\omega=m$ to $\omega=m+\Delta$,
integration of (\ref{Fapprox}) gives the estimate
\begin{equation}
\label{Fbar} \bar F\equiv  \frac {1}{\Delta} \int_{m}^{m+\Delta}
d\omega F(\omega) \simeq  \frac {\sqrt{2}\beta^2} {\Delta^{1/2}
m^{3/2}}\sin^2(\frac{1}{2}mL) ~~~.
\end{equation}
The formulas (\ref{Pbar}) and (\ref{Fbar}) will be of use for
estimating the magnitude of photon-axion and axion-photon
conversion effects in the threshold region.

\section{Magnetic field penetrating the wall}

Up to this point we have assumed that the magnetic field is
present in a vacuum region, where there is no photon absorption.
Let us now briefly examine the situation where the magnetic field
and the photon absorbing ``wall'' overlap.  In this case the
photon propagation equation must include a complex dielectric
constant $n=n_R+in_I$ with nonzero imaginary part $n_I$, and so
the eigenmode equations (\ref{modeqs}) are modified to read
\begin{eqnarray}
\label{nmodeqs} (-\omega^2+K^2  + m^2)\phi &=&
-i\omega \beta a~~~, \nonumber\\
 (-n^2\omega^2+K^2) a & = & i\omega \beta \phi~~~,
\end{eqnarray}
which now require $K$ to obey the quartic equation
\begin{equation}
\label{nKeq}
(-\omega^2+K^2+m^2)(-n^2\omega^2+K^2)=\omega^2\beta^2~~~.
\end{equation}
For $\beta/m<<1$, the two types of eignmodes are a ``photon-like''
mode, where $-n^2\omega^2+K^2$ is small, so that
\begin{equation}
\label{photonlike} -n^2\omega^2+K^2= \frac {\omega^2
\beta^2}{-\omega^2+K^2+m^2} \simeq \frac {\omega^2 \beta^2} {m^2
+(n^2-1)\omega^2}~~~,
\end{equation}
and an ``axion-like'' mode, where $-\omega^2+K^2+m^2$ is small, so
that
\begin{equation}
\label{axionlike} -\omega^2+K^2+m^2=\frac{\omega^2\beta^2}
{-n^2\omega^2+K^2}\simeq  - \frac{\omega^2\beta^2} {m^2
+(n^2-1)\omega^2}~~~.
\end{equation}
Thus,
\begin{eqnarray}
\label{kvalues}
 K_{\rm photon} &\simeq& n_R \omega +i
n_I\omega~~~, \nonumber
\\
K_{\rm axion} &\simeq& \sqrt{\omega^2-m^2} + i \frac
{n_Rn_I}{\sqrt{\omega^2-m^2} } \frac {\omega^4 \beta^2}
 {|m^2+(n^2-1) \omega^2|^2}~~~.
\end{eqnarray}

We see that for $\omega >> m$ and $n_I\omega>0$,  the
``photon-like'' mode decays as $e^{-\sigma_{\rm photon} z}$, with
$\sigma_{\rm photon}\simeq n_I \omega$, while the ``axion-like''
mode decays much more weakly, as $e^{-\sigma_{\rm axion}z}$, with
\begin{equation}
\label{sigmarat} \sigma_{\rm axion} \simeq \sigma_{\rm photon}
\frac {n_R \beta^2} {|n^2-1|^2 \omega^2}~~~.
\end{equation}
Putting in typical numbers, $\omega \sim 1 {\rm eV}$, $\beta  \sim
10^{-11} {\rm eV}$, and $\sigma_{\rm photon} \sim 10^5 {\rm
cm}^{-1}$, one has $\sigma_{\rm axion} \sim 10^{-17} {\rm
cm}^{-1}$, and hence there is negligible decay of the
``axion-like'' wave over a one meter flight path.

\section{Summary and discussion}

To summarize, we have given three different methods for
calculating photon-axion and axion-photon conversion in a magnetic
field.  To get the lowest order transmission amplitude, the WKB
method is clearly the simplest, since it involves only the
integration of a first order differential equation.  To also get
the reflected wave in leading order, the WKB/eikonal and Green
function methods are of about equal complexity.  To get an all
orders answer, a wave matching calculation involving $4 \times 4$
matrices is necessary. Although this calculation was done only for
piecewise constant magnetic fields, by expanding the formulas of
Appendix B for infinitesimal $L$ as $S(\beta,dL)=1+dL G(\beta)$,
and using the group property
\begin{equation}
\label{group1} S\big(\beta(z\leq L+dL),L+dL\big)=S\big(\beta(z\leq
L), L\big)S\big(\beta(L), dL\big) =S\big(\beta(z\leq L),
L\big)[1+dL G\big(\beta(L)\big)]~~~,
\end{equation}
one gets a differential equation
\begin{equation}
\label{group2} \frac {dS\big(\beta(z\leq L),L\big)}
{dL}=S\big(\beta(z\leq L),L\big) G\big(\beta(L)\big)~~~,
\end{equation}
which can be integrated to give the all orders $S$ for an
arbitrary longitudinal magnetic field profile $\beta(z)$.

All three calculation methods show that when the axion mass is not
set equal to zero, there is an enhancement of the photon-axion
conversion amplitude near threshold at $\omega=m$.  The lowest
order calculations indicate a threshold cusp that violates
unitarity for small enough $k$, and so the all orders calculation
is needed to see how, for $k$ very close to threshold, unitarity
is restored. Deciding whether  this threshold enhancement can be
used for new experimental searches for axions will require further
careful analysis.  For example, when $\omega >> m$ the light
through walls probability obtained from (\ref{transmodsq}) is
$P(\omega>>m)=(\beta L/2)^4$, which for large values of the
effective path length $mL$ is much larger than the probability
(\ref{Pbar}) for the same geometry when $\omega$ is near
threshold, assuming that the effective bandwidth $\Delta$ is of
order the axion mass $m$.  However, further investigation will be
needed to see if (\ref{Pbar}) suggests ways of exploiting the
threshold enhancement by using very small laser bandwidths $\Delta
<< m$.   Similarly, further  analysis will be needed to determine
whether the enhanced axion flux near threshold given by
(\ref{Fbar}) has astrophysical implications in strong magnetic
field environments.

\section{Acknowledgments}

The authors wish to thank E. I. Guendelman for his participation
in initial phases of this project, and for helpful email
correspondence suggesting a WKB approach and pointing out the role
of the $k$ factor in the unitarity relation.   The work of  SLA
was supported in part by the U. S. Department of Energy under
Grant No. DE-FG02-90ER40542, JG and FM were partially supported by grants from FONDECYT 1050114 and  1060079.
JL-S acknowledges the support from MEC/FULBRIGHT-FU2006-0469.

\section{Added Note}

Carlo Rizzo has raised the pertinent question of how the 
factor $\sin^2(\frac {1}{2} mL)$, and its square,  in the formulas 
(103)--(106), are to be interpreted when $mL>>1$.  This factor is the 
evaluation at threshold of the factor $\sin^2\big(\frac {1}{2} L 
(\omega-\ka)\big)\simeq \sin^2\big(\frac{1}{2}L (m-\sqrt{2 m (\omega-m)})\big)$.
When $\frac{1}{2}L\sqrt{2 m \Delta} >>1$, the sine function has many oscillations 
over the integration interval in (106), (104), and it then  can be 
replaced by the respective averages $\langle \sin^2\big(\frac{1}{2}L (m-\sqrt{2 m (\omega-m)})\big) \rangle_{AV} =1/2$, 
and $\langle \sin^4\big(\frac{1}{2}L (m-\sqrt{2 m (\omega-m)})\big) \rangle_{AV} =3/8$, in (106) and (104) respectively. 
When $\frac{1}{2}L\sqrt{2 m \Delta}$ is of order unity, the full $\omega$-dependent 
form of the argument of the sine function
 should be included in the integrals in (106), (104) in place of 
its threshold evaluation, and the integrals can then be computed  
numerically.

\appendix\section{Matrices for matching at $z=0$ and
$z=L$}

The matrices $M_1$, $M_2$, and $M_{12}=M_1^{-1}M_2$ needed for the
$z=0$ match are
\begin{eqnarray}
\label{tranzero1} M_1=&& \left(
\begin{array}{cccc}
1 & 1 & 0 & 0
\\
0 & 0 & 1 & 1
\\
k &-k & 0 & 0
\\
0 & 0 & \omega & -\omega
\end{array}\right)~~~,
\nonumber
\\
M_2=&&\left(
\begin{array}{cccc}
1 & 1 & \delta & \delta
\\
\delta & \delta & 1 & 1
\\
k_- & -k_- & \delta~k_+ & -\delta~k_+
\\
\delta~k_-& -\delta~k_- & k_+ & -k_+
\end{array}\right)~~~,
\nonumber
\\
M_{12}=&&\frac{1}{2}\left(
\begin{array}{cccc}
1+{k_-}/{k}&1-{k_-}/{k}&\delta\,(1+{k_+}/{k})&\delta\,(1-{k_+}/{k})
\\
1-{k_-}/{k}&1+{k_-}/{k}&\delta\,(1-{k_+}/{k})&\delta\,(1+{k_+}/{k})
\\
\delta\,(1+{k_-}/{\omega})&\delta\,(1-{k_-}/{\omega})&1+{k_+}/{\omega}&1-{k_+}/{\omega}
\\
\delta\,(1-{k_-}/{\omega})&\delta\,(1+{k_-}/{\omega})&1-{k_+}/{\omega}&1+{k_+}/{\omega}
\end{array}\right)~~~.
\end{eqnarray}

The matrices $M_3$, $M_4$, and $M_{34}=M_3^{-1}M_4$ needed for the
$z=L$ match are

\begin{eqnarray}
\label{tranzL} M_3=&&\left(
\begin{array}{cccc}
e^{iLk_-} &  e^{-iLk_-} & \delta\,e^{iLk_+}  & \delta\,e^{-iLk_+}
\\
\delta\,e^{iLk_-} & \delta\,e^{-iLk_-} & \,e^{iLk_+} &
\,e^{-iLk_+}
\\
k_-\,e^{iLk_-} & -k_-\,e^{-iLk_-}& \delta~k_+ \,e^{iLk_+}&
-\delta~k_+\,e^{-iLk_+}
\\
\delta~k_-\,e^{iLk_-}& -\delta~k_- \,e^{-iLk_-}& k_+\,e^{iLk_+} &
-k_+\,e^{-iLk_+}
\end{array}\right)~~~,
\nonumber
 \\
M_4=&&\left(
\begin{array}{cccc}
e^{iLk} & e^{-iLk}& 0 &0
\\
0 & 0 & e^{iL \omega} & e^{-iL\omega}
\\
k\,e^{iLk} & -k\,e^{-iLk} & 0 & 0
\\
0 & 0 & \omega\,e^{iL\omega} & -\omega\,e^{-iL\omega}
\end{array}\right)~~~,
\nonumber
\\
M_{34}=&&\frac {1}{2(1-\delta^2)}\nonumber\\ \times &&\left(
\begin{array}{cccc}
(1+k/k_-)e^{iL(k-k_-)}&(1-k/k_-)e^{-iL(k+k_-)}&-\delta\,(1+\omega/k_-)e^{iL(\omega-k_-)}
&-\delta\,(1-\omega/k_-)e^{-iL(\omega+k_-)}
\\
(1-k/k_-)e^{iL(k+k_-)}&(1+k/k_-)e^{iL(k_--k)}&-\delta\,(1-\omega/k_-)e^{iL(\omega+k_-)}
&-\delta\,(1+\omega/k_-)e^{iL(k_--\omega)}
\\
-\delta\,(1+k/k_+)e^{iL(k-k_+)}&-\delta\,(1-k/k_+)e^{-iL(k+k_+)}&(1+\omega/k_+)e^{iL(\omega-k_+)}
&(1-\omega/k_+)e^{-iL(\omega+k_+)}
\\
-\delta\,(1-k/k_+)e^{iL(k+k_+)}&-\delta\,(1+k/k_+)e^{iL(k_+-k)}&(1-\omega/k_+)e^{iL(\omega+k_+)}
&(1+\omega/k_+)e^{iL(k_+-\omega)}
\end{array}\right)~~~.\nonumber
\\
\end{eqnarray}

\section{Matrix elements of $S_{ij}$}
\begin{eqnarray}
2 \left(\delta ^2-1\right) k_- k_+    k ~{ S_{11}} &=&  e^{i L k}
\bigg[ik_+(k_-^2+k^2)\sin(Lk_-)-2k_+k_-k\cos(Lk_-)\nonumber
\\
&& -\delta^2~[ik_-(k_+^2+k^2)\sin(Lk_+) -2k_+\,k_-\,k\cos(Lk_+)]
\bigg]
\\
2 \left(\delta ^2-1\right) k_- k_+    k ~{ S_{12}} &=& i e^{-i L
k} \bigg[
k_+(k_-^2-k^2)\sin(Lk_-)+\delta^2\,k_-(k^2-k_+^2)\sin(Lk_+)\bigg]
\\
 2\left(\delta ^2-1\right) k_- k_+ k~{ S_{13}}
&=& \delta\,e^{iL\omega}
\bigg[(\omega+k)k_-k_+[\cos(Lk_-)-\cos(Lk_+)]\nonumber
\\
&& -i[k_+(k_-^2+\omega k)\sin(Lk_-) - k_-(k_+^2+\omega
k)\sin(Lk_+)] \bigg]
\\
2\left(\delta ^2-1\right) k_- k_+ k~{ S_{14}} &=&
\delta\,e^{-iL\omega}
\bigg[(\omega-k)k_-k_+[\cos(Lk_+)-\cos(Lk_-)]\nonumber
\\
&& -i[k_+(k_-^2 - \omega k)\sin(Lk_-) - k_-(k_+^2 - \omega
k)\sin(L k_+)] \bigg]
\\
{ S_{21}} &=&-e^{2iLk}~{ S_{12}}
\\
2(\delta^2-1)~k_- k_+ k~{ S_{22}} &=&e^{-iLk}\bigg[
-k_+[2k_-k\cos(Lk_-) +i(k_-^2+k^2)\sin(Lk_-)] \nonumber
\\
&&+\delta^2k_-[2k_+k\cos(Lk_+) +i(k_+^2 + k^2)\sin(Lk_+)] \bigg]
\\
2(\delta^2-1)k_-\,k_+\,k\,{ S_{23}} &=& -e^{i L \omega } \delta
\bigg[k_+[ k_-  \left(\omega -k\right)\cos
   \left(L k_-\right)  -i \left(k_-^2-\omega  k\right)\sin \left(L k_-\right) ] \nonumber
\\
&&-k_- [ k_+ \left(\omega-k \right)\cos \left(L
   k_+\right)-i \left(k_+^2-\omega
   k\right)\sin \left(L
   k_+\right) ]
\bigg]
\\
2\left(\delta ^2-1\right) k_- k_+ k~ { S_{24}} &=&
e^{-iL\omega}\delta\bigg[-k_-[k_+(\omega+k)\cos(Lk_+) +i(k_+^2
+\omega\,k)\sin(Lk_+)] \nonumber
\\
&& +k_+[k_-(\omega+k)\cos(Lk_-)+i(k_-^2+\omega\,k)\sin(Lk_-)]
\bigg]
\\
{ S_{31}}&=&-e^{-i L \left(\omega -k\right)}~ \frac{k}{\omega }~{
S_{13}}
\\
{ S_{32}}&=&e^{-i L \left(\omega +k\right)}~ \frac{k}{\omega }~{
S_{23}}
\\
2 \left(\delta ^2-1\right) \omega  k_- k_+{ S_{33}}&=& i\,e^{i L
\omega }\bigg[k_-[ 2i\omega k_+\,\,\cos(Lk_+)+ (\omega^2
+k_+^2)\sin(Lk_+)]\nonumber
\\
&&
-\delta^2\,k_+[2i\omega\,k_-\cos(Lk_-)+(\omega^2+k_-^2)\sin(Lk_-)
] \bigg]
\\
2(\delta^2-1)\omega  k_- k_+\,{ S_{34}} &=& ie^{-iL\omega}\bigg[
k_-(k_+^2 -\omega^2)\sin(Lk_+) -
\delta^2\,k_+(k_-^2-\omega^2)\sin(Lk_-) \bigg]
\\
{ S_{41}} &=& e^{iL(\omega +k)}~\frac{k}{\omega}~{ S_{14}}
\\
{ S_{42}} &=& -e^{iL(\omega - k)}~\frac{k}{\omega}~{ S_{24}}
\\
{ S_{43}} &=& -e^{2iL\omega}~{ S_{34}}
\\
2\left(\delta ^2-1\right) \omega  k_- k_+{ S_{44}}&=&e^{-iL\omega}
\bigg[-k_-[2\omega\,k_+\,\cos(Lk_+)+i(\omega^2+k_+^2)\sin(Lk_+)]
\nonumber
\\
&&+\delta^2k_+[2\omega\,k_-\,\cos(Lk_-)+i(\omega^2+k_-^2)\sin(Lk_-)]
\bigg]
\\
\nonumber
\end{eqnarray}
The matrix elements of ${ S}$ have a large number of symmetries
that arise from the structure of $M_{12}$ and $M_{34}$.  First,
all  of the ${ S}_{ij}$ are even functions of  $k_+$ and even
functions of $k_-$.  Second, the matrix elements above have one of
the following two structures: (i)~ $[\delta/(1-\delta^2)]
[f(k_+,k_-,k,\omega)-f(k_-,k_+,k,\omega)]$,~~(ii)~
$[1/(1-\delta^2)] [\delta^2
f(k_+,k_-,k,\omega)-f(k_-,k_+,k,\omega)]$, with the functions $f$
different for each matrix element.  Since we have seen that under
the interchange $k_+ \leftrightarrow k_-$ one has $\delta
\leftrightarrow 1/\delta$, these two structures are both invariant
under the interchange of $k_+$ and $k_-$, and so all matrix
elements ${ S}_{ij}$ are invariant under this interchange.
Finally, there are a number of relations between matrix elements
under reversal of the sign of $k$ or $\omega$, as follows
\begin{eqnarray}
\label{symms} &&{ S}_{11}={ S}_{22}|_{k\to -k}~,~~ { S}_{21}={
S}_{12}|_{k\to -k}~,~~ { S}_{32}={ S}_{31}|_{k\to -k}~,~~{
S}_{13}={ S}_{23}|_{k\to -k}~,~~{ S}_{41}={ S}_{42}|_{k\to
-k}~,~~{ S}_{14}={ S}_{24}|_{k\to -k}~,~~\nonumber
\\
&&{ S}_{33}={ S}_{44}|_{\omega\to -\omega}~,~~{ S}_{34}={
S}_{43}|_{\omega\to -\omega}~,~~{ S}_{13}={ S}_{14}|_{\omega\to
-\omega}~,~~{ S}_{23}={ S}_{24}|_{\omega\to -\omega}~,~~{
S}_{41}={ S}_{31}|_{\omega\to -\omega}~,~~{ S}_{42}={
S}_{32}|_{\omega\to -\omega}~~~.\nonumber
\\
\end{eqnarray}
These relations serve as useful checks on the calculation.


\begin{figure}[ht!]
\begin{center}
\includegraphics[width=.50\textwidth]{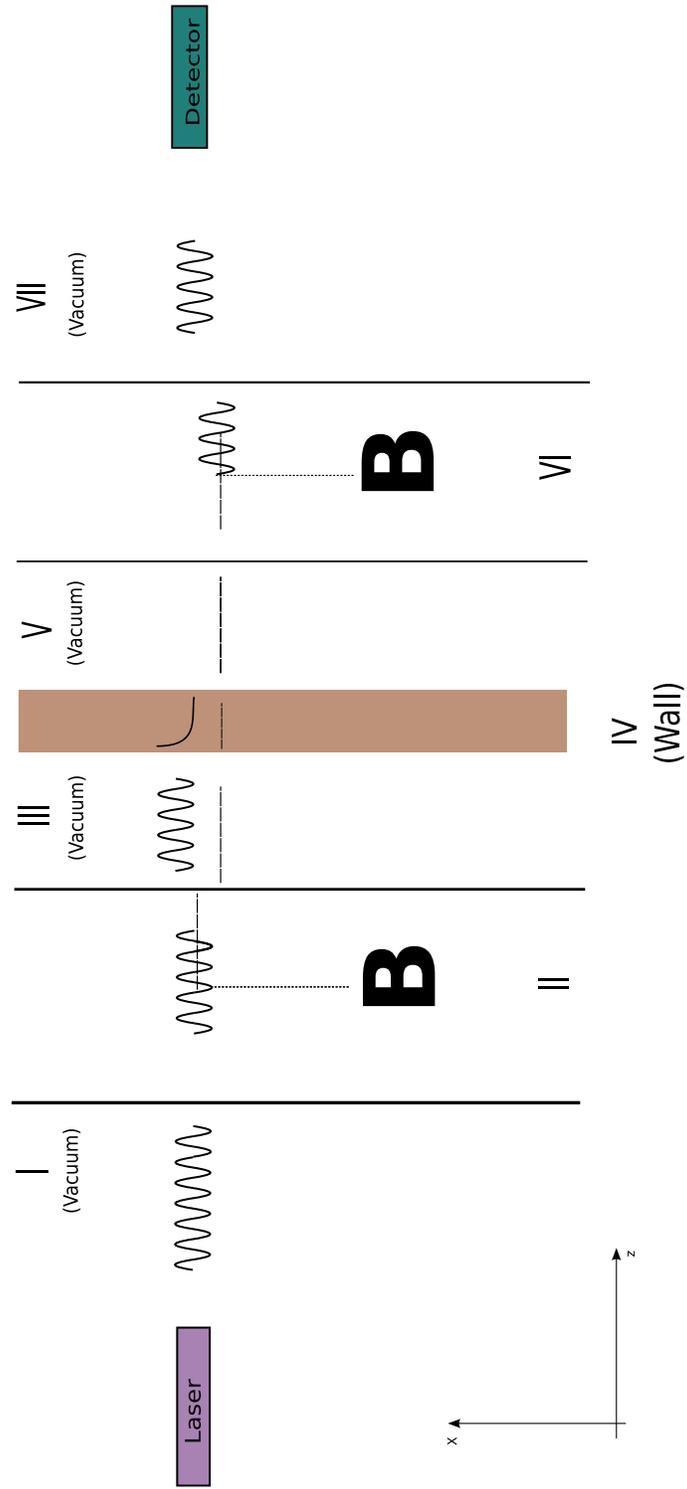}\
\caption{Light shining through a wall setup.  The wavy line
indicates the photon field and the dashed line the axion field. In
region I, there is only the photon wave, in region II the magnetic
field gives rise to an axion wave, which exits with the photon
wave into region III.  In region IV, the ``wall'', the photon wave
is absorbed, leaving only the axion wave in region V. In region
VI, the magnetic field regenerates a photon wave from the axion,
which exits into region VII, where the regenerated photon is
detected. }\label{fig}
\end{center}
\end{figure}


\end{document}